\documentclass[aps,prb,preprint,superscriptaddress]{revtex4-1}
\usepackage{graphicx}
\usepackage{amsmath}
\begin{document}

%\preprint{}

\title{Intra-layer doping effects on the high-energy magnetic correlations in NaFeAs}



\author{Jonathan Pelliciari}
\email[]{jonathan.pelliciari@psi.ch}
\affiliation{Department of Synchrotron Radiation and Nanotechnology, Paul Scherrer Institut, CH-5232 Villigen PSI, Switzerland}
\author{Yaobo Huang}
\affiliation{Department of Synchrotron Radiation and Nanotechnology, Paul Scherrer Institut, CH-5232 Villigen PSI, Switzerland}
\affiliation{Beijing National Lab for Condensed Matter Physics, Institute of Physics, Chinese Academy of Sciences P. O. Box 603, Beijing 100190, China}
\author{Tanmoy Das}
\affiliation{Department of Physics, Indian Institute of Science, Bangalore-560012, India}   
\author{Marcus Dantz}
\affiliation{Department of Synchrotron Radiation and Nanotechnology, Paul Scherrer Institut, CH-5232 Villigen PSI, Switzerland}
\author{Valentina Bisogni}
\affiliation{Department of Synchrotron Radiation and Nanotechnology, Paul Scherrer Institut, CH-5232 Villigen PSI, Switzerland}
\affiliation{National Synchrotron Light Source II, Brookhaven National Laboratory, Upton, NY 11973, USA}
\author{Paul Olalde Velasco}
\affiliation{Department of Synchrotron Radiation and Nanotechnology, Paul Scherrer Institut, CH-5232 Villigen PSI, Switzerland}
\author{Vladimir N. Strocov}
\affiliation{Department of Synchrotron Radiation and Nanotechnology, Paul Scherrer Institut, CH-5232 Villigen PSI, Switzerland}
\author{Lingyi Xing}
\affiliation{Beijing National Lab for Condensed Matter Physics, Institute of Physics, Chinese Academy of Sciences P. O. Box 603, Beijing 100190, China}
\author{Xiancheng Wang}
\affiliation{Beijing National Lab for Condensed Matter Physics, Institute of Physics, Chinese Academy of Sciences P. O. Box 603, Beijing 100190, China}
\author{Chuangqing Jin}
\affiliation{Beijing National Lab for Condensed Matter Physics, Institute of Physics, Chinese Academy of Sciences P. O. Box 603, Beijing 100190, China}
\affiliation{Collaborative Innovation Center for Quantum Matters, Beijing, China}
\author{Thorsten Schmitt}
\email[]{thorsten.schmitt@psi.ch}
\affiliation{Department of Synchrotron Radiation and Nanotechnology, Paul Scherrer Institut, CH-5232 Villigen PSI, Switzerland}

\date{\today}

\begin{abstract}
We have used Resonant Inelastic X-ray Scattering (RIXS) and dynamical susceptibility calculations to study the magnetic excitations in NaFe$_{1-x}$Co$_x$As (x = 0, 0.03 and 0.08). Despite a relatively low ordered magnetic moment, collective magnetic modes are observed in parent compounds (x = 0) and persist in optimally (x = 0.03) and overdoped (x = 0.08) samples. Their magnetic bandwidths are unaffected by doping within the range investigated. High energy magnetic excitations in iron pnictides are robust against doping, and present irrespectively of the ordered magnetic moment. Nevertheless, Co doping slightly reduces the overall magnetic spectral weight, differently from previous studies on hole-doped BaFe$_{2}$As$_{2}$, where it was observed constant. Finally, we demonstrate that the doping evolution of magnetic modes is different for the dopants being inside or outside the Fe-As layer. 
\end{abstract}

%\keywords{}

\maketitle

\section{\label{sec:level1}Introduction}

Superconductivity (SC) in iron pnictides (FePns) emerges after doping a metallic antiferromagnet \cite{kamihara_iron-based_2008}. Proximity, competition and even coexistence between antiferromagnetic ordering (AF) and SC have been observed in the phase diagram in many FePns systems, with others, such as LiFeAs and KFe$_{2}$As$_{2}$, not showing any ordering in the SC dome. Similar behavior is also seen in cuprates and heavy-fermion superconductors \cite{scalapino_common_2012, chubukov_pairing_2012}, leading to the unsolved puzzle on whether a static magnetic order is a prerequisite to unconventional SC, or dynamical spin fluctuations without preferential wavevector are essential for superconducting pairing, irrespective of magnetic order. Therefore, the experimental characterization of dynamical spin fluctuations is vital to confirm or dismantle such concepts. In this context, Inelastic Neutron Scattering (INS) is at the forefront of research in the study of magnetic excitations in unconventional superconductors (see Refs. \cite{tranquada_superconductivity_2014, dai_antiferromagnetic_2015, inosov_spin_????, fujita_progress_2011}, and references therein). However, momentum resolved experimental detection of magnetic modes above 90 meV still represents a challenging task. This has been overcome by Resonant Inelastic X-ray Scattering (RIXS) employing resonances at absorption edges (such as Fe-L$_{2, 3}$ or Cu-L$_{2, 3}$) which increase the cross section by several orders of magnitudes compared to INS. Recently, RIXS has been used to successfully detect high energy magnetic excitations in superconducting cuprates \cite{dean_spin_2012, le_tacon_intense_2011, lee_asymmetry_2014, ishii_high-energy_2014,  peng_magnetic_2015, minola_collective_2015, braicovich_magnetic_2010} and FePns \cite{zhou_persistent_2013}. Moreover, INS and RIXS techniques complement each other in their span of the reciprocal space, with RIXS probing around the $\Gamma$ point and INS  measuring around the antiferromagnetic ordering vector point in the Brillouin zone \cite{dean_insights_2015,  dai_antiferromagnetic_2015, tranquada_superconductivity_2014, ament_resonant_2011}.

Although the exact origin of magnetism in FePns has not been fully established, they usually present a sizable ordered magnetic moment ($\mu$) in the order of $\approx$1 $\mu_{B}$ \cite{johnston_puzzle_2010, stewart_superconductivity_2011}. NaFeAs contrasts with other FePns because of its low $\mu$ (about 0.1 $\mu_{B}$) which differs in value with the most studied BaFe$_{2}$As$_{2}$ ($\approx$1.1-1.3 $\mu_{B}$) by an order of magnitude \cite{johnston_puzzle_2010, stewart_superconductivity_2011}.
Nonetheless, NaFeAs shows AF ordering even though at lower T$_{N}$ (ca. 45 K vs ca. 140 K for NaFeAs and BaFe$_{2}$As$_{2}$, respectively) \cite{li_structural_2009, johnston_puzzle_2010}. The source of such change in T$_{N}$  has been ascribed to the larger distance between the As and the Fe layer in NaFeAs that increases electronic correlations consequently affecting also the spin wave spectrum \cite{yin_kinetic_2011, zhang_effect_2014}.
When Co atoms substitute Fe within the Fe-As layer, AF is suppressed and SC arises with a T$_C\approx$ 20 K at optimal doping \cite{stewart_superconductivity_2011}. The significant increase of both quasiparticle scattering rate and bandwidth as a function of Co substitution, as seen by angle resolved photoelectron spectroscopy, have been ascribed to the decrease of the Fe-As bond length due to doping. As a consequence Hund$\textsc{\char13}$s coupling and electronic correlations decrease upon Co doping \cite{ye_extraordinary_2014}. This, in turn, should affect the behavior of spin fluctuations likely connected with superconductivity \cite{scalapino_common_2012, chubukov_pairing_2012}.

Here, we report on the measurements of high energy magnetic fluctuations in NaFe$_{1-x}$Co$_{x}$As samples with x = 0.0 (parent compound, Par), x = 0.03 (close to optimal doping, T$_c$ = 20 K, Opt) and x = 0.08 (overdoped, T$_c$ = 6 K, Over) by means of RIXS. The bandwidth of these spin excitations is similar for Par, Opt and Over. The presence of magnetic modes along (1, 0) and (1, 1) reciprocal directions in all samples is striking, displaying the coexistence of these modes with SC and their persistence even for overdoped samples. The nature of these excitations has been compared to self-consistent calculations of dynamical spin ($\chi_{S}$) and charge ($\chi_{C}$) susceptibilities by including the self-energy correction due to coupling of these modes to the electronic structure \cite{das_intermediate_2014}. Our calculations show the dominance of $\chi_{S}$ over $\chi_{C}$. The experimental spectral weight of the magnetic modes is qualitatively observed to decrease when Co doping is increased.

\begin{figure}
\includegraphics[scale=0.4]{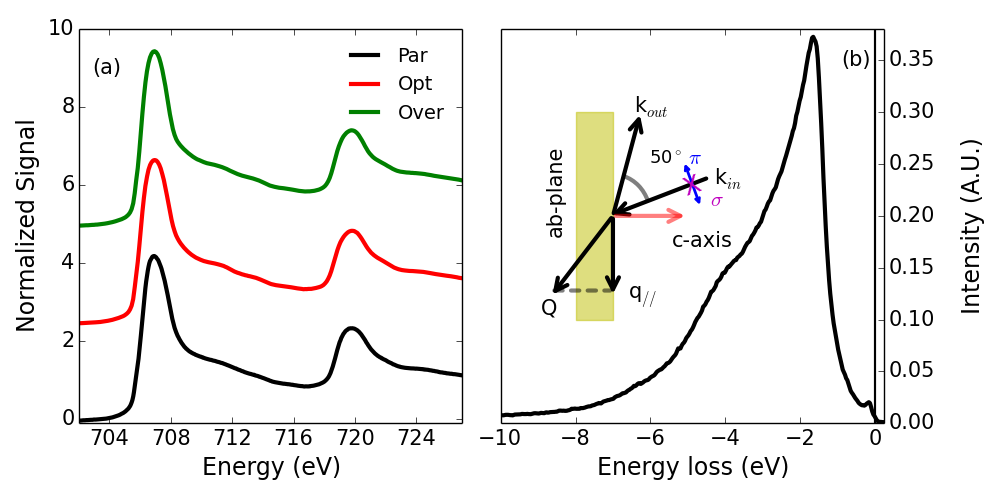}
\caption{\label{fig:fig1} (a) Fe L$_{2, 3}$ X-Ray Absorption Spectra for Par (black line), Opt (red line) and Over (green line). (b) Experimental configuration and Fe L$_3$ RIXS spectra for Par at (0.44, 0) and $h\nu_{in}$=707.3 eV. }
\end{figure}

\section{\label{sec:level1}Experimental and methods}
\subsection{\label{sec:level2}Samples preparation}
Single crystals of NaFe$_{1-x}$Co$_{x}$As were grown by the self-flux method, using NaAs as the flux. The precursor Na$_{3}$As was obtained by mixing Na lump and As powder, which were then sealed in an evacuated titanium tube and sintered at $650\,^{\circ}\mathrm{C}$ for 10 h. Fe$_{1-x}$Co$_{x}$As precursors were prepared by mixing Fe, Co and As powder thoroughly, pressed into pellets, sealed in a evacuated quartz tube, before being sintered at $700\,^{\circ}\mathrm{C}$ for 20h. To ensure the homogeneity of the product, these pellets were grounded and sintered another time. The stoichiometric amount of Na$_{3}$As, Fe$_{1-x}$Co$_{x}$As and As powder were weighed according to the element ratio of Na(Fe$_{1-x}$Co$_{x}$)$_{0.3}$As. The mixture was grounded and put into an alumina crucible and sealed in a Nb crucible under 1 atm of Argon gas. The Nb crucible was then sealed in an evacuated quartz tube and heated to $900\,^{\circ}\mathrm{C}$ before being slowly cooled down to $600\,^{\circ}\mathrm{C}$ ($3\,^{\circ}\mathrm{C}$/h) to grow single crystals. All sample preparations, except for sealing, were carried out in a glove box filled with high purity Argon gas. The element composition of the NaFe$_{1-x}$Co$_{x}$As single crystals was checked by energy dispersive x-ray spectroscopy (EDS). Samples were stored in a sealed quartz tube and prepared for spectroscopic studies in a glove box under high purity N$_2$ flow to avoid contact with air. 

\subsection{\label{sec:level2}Experimental conditions}
The samples were mounted with the \textit{ab} plane perpendicular to the scattering plane and the \textit{c} axis lying in it (sketch in Fig.\ref{fig:fig1}b) and post-cleaved in situ at a pressure better than 2.0x10$^{-10}$ mbar. The reciprocal space directions studied are (1, 0) and (1, 1) according to the orthorhombic unfolded crystallographic notation \cite{park_symmetry_2010}. The values of momentum transferred are expressed as relative lattice units (R.L.U.) (q$_{//}\cdot$a/2$\pi$). We use the convention of 1 Fe per unit cell. All the measurements were carried out at 13 K by cooling the manipulator with liquid helium.
X-ray Absorption Spectra (XAS) and RIXS experiments were performed at the ADRESS beamline of the Swiss Light Source, Paul Scherrer Institute, Villigen PSI, Switzerland \cite{strocov_high-resolution_2010, ghiringhelli_saxes_2006}. XAS spectra were measured in Total Fluorescence Yield (TFY). We measured Fe L$_{2, 3}$ XAS spectra for all samples at 65$^{\circ}$ incidence angle relative to the sample surface. The RIXS spectrometer was set to a scattering angle of 130$^{\circ}$ and the incidence angle on the samples surface was varied to change the in plane momentum transferred (q) from (0, 0) to (0.44, 0) and from (0, 0) to (0.32, 0.32) as shown in the sketch of Fig.\ref{fig:fig1}b. All measurements in the present paper are recorded in grazing emission configuration (Fig.\ref{fig:fig1}b). The zero energy loss in our RIXS spectra has been estimated measuring spectra in $\sigma$ polarization. The total energy resolution has been measured employing the elastic scattering of carbon-filled acrylic tape and is around 110 meV. 

\subsection{\label{sec:level2}Calculations}
We performed dynamical spin ($\chi_{S}$) and charge ($\chi_{C}$) susceptibilities calculations. The calculations self-consistently include the many-body corrections within a random-phase approximation (RPA) as well as the self-energy correction due to both the spin and charge fluctuations starting from the DFT band structure \cite{das_intermediate_2014}. The calculation includes intra- and inter- orbital Hubbard U, as well as Hund$\textsc{\char13}$s coupling and pair-hopping interactions \cite{Uvalues}. The doping effect is modeled by rigid band shift. The bare values of both $\chi_{S}$ and $\chi_{C}$, without including the self-energy correction, overestimate the energy scales, and underestimate the weights of the collective excitations. The inclusion of the self-energy due to the coupling of the spin and charge fluctuations to the electronic state renormalizes the quasiparticle bands by about 30\%, in accordance with various photoemission data \cite{yi_electronic_2012, ye_extraordinary_2014}. This self-consistent approach has been successfully applied to describe the role of renormalized spin and charge excitations on superconductivity in cuprates \cite{das_intermediate_2014}, heavy-fermions \cite{das_spin_2012} and transition-metal dichalcogenide superconductors \cite{das_superconducting_2015}.

\begin{figure}
\includegraphics[scale=0.4]{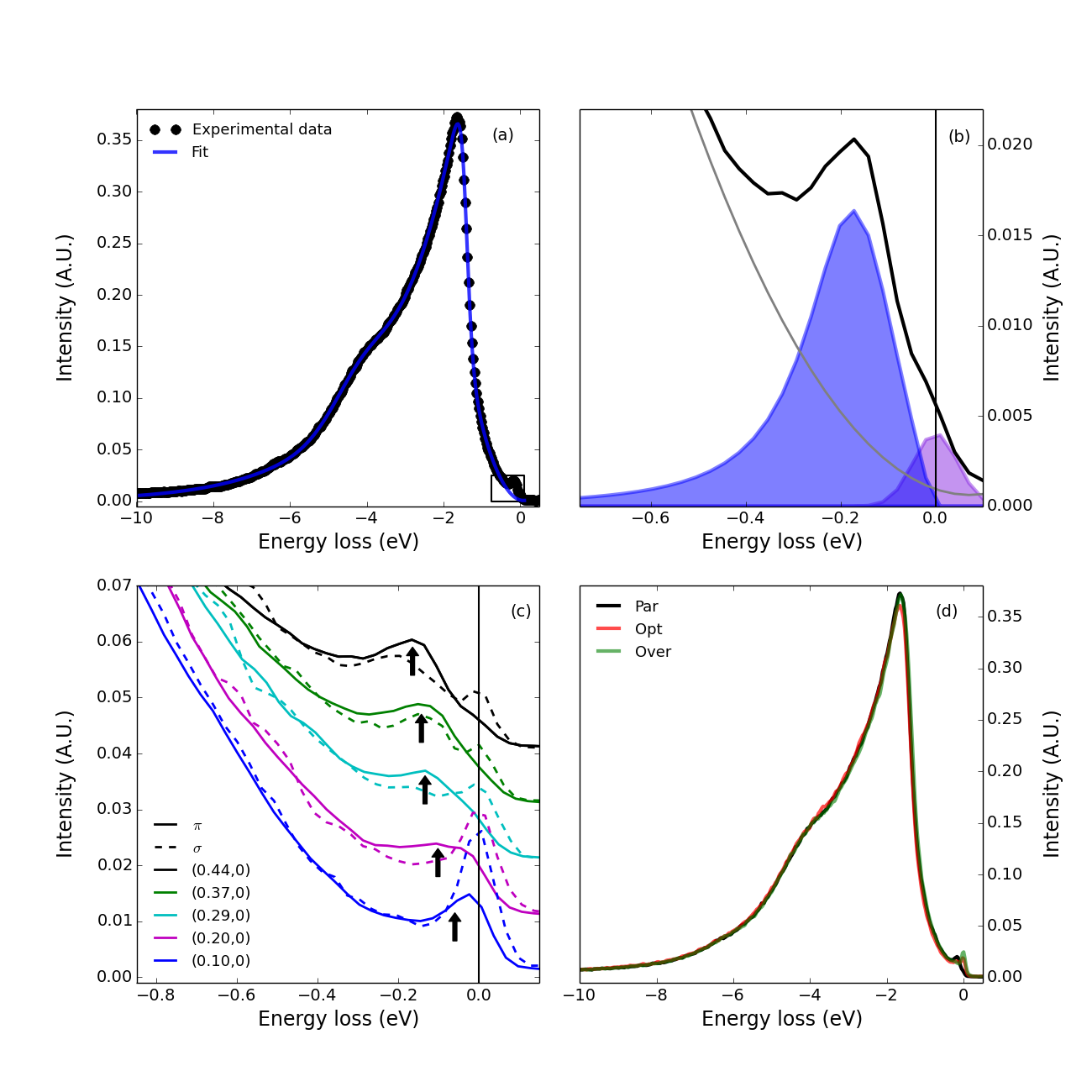}
\caption{\label{fig:fig2} (a) Experimental data (black dots) and relative fitting (blue solid line) of resonant emission of Par at (0.44, 0) and $\pi$  polarization. The incoming energy was tuned on the maximum of the Fe L$_3$ XAS.
(b) Zoom into the low energy loss region of (a) and fitting of background (gray solid line), elastic (pink shaded) and magnetic peak (purple shaded). (c) Momentum and polarization dependence of the RIXS spectra of Par along (0,0)$\rightarrow$(0.44, 0). Solid line is $\pi$ polarization and dashed line is $\sigma$ polarization. (d) Comparison of RIXS spectra at (0.44, 0) and $\pi$  polarization for Par, Opt and Over. All the data were collected at 13 K.}
\end{figure}

\section{\label{sec:level1}Results and discussion}
 Fig.\ref{fig:fig1}a displays the Fe-L$_{2, 3}$ XAS spectra of Par, Opt and Over, respectively collected at 65$^{\circ}$ incidence angle. The spectra are composed of a broad peak centered at 707 eV, typical of metallic systems containing iron \cite{zhou_persistent_2013, kurmaev_identifying_2009, yang_evidence_2009}. The incident energy for RIXS was tuned at the main Fe-L$_3$ peak. In Fig.\ref{fig:fig1}b an exemplary RIXS spectrum of Par at q = (0.44, 0) is shown. The main line in this spectrum resembles emission from metallic systems, with a broad asymmetric peak displaying a maximum at around -2 eV in energy loss ($h\nu_{out}$-$h\nu_{in}$) arising from resonant emission of itinerant electrons \cite{zhou_persistent_2013, kurmaev_identifying_2009, hancock_evidence_2010}. Normalization of the spectra is carried out integrating the area between -0.4 eV and -10.0 eV. This spectral region refers directly to the amount of Fe in the samples due to the element sensitivity of RIXS allowing direct comparison of samples despite different Fe content.

\begin{figure}
\includegraphics[width=\textwidth]{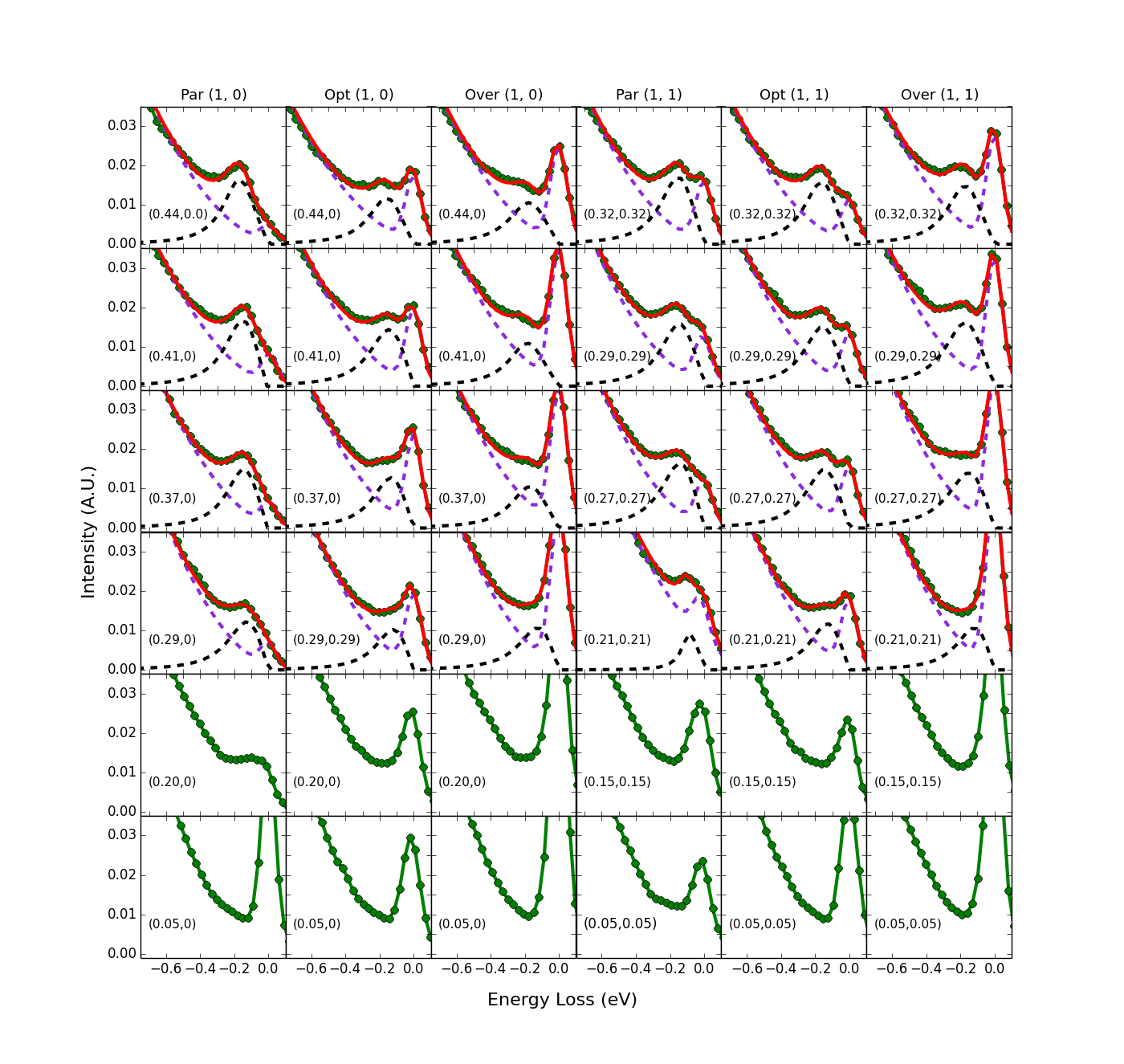}
\caption{\label{fig:fig3} Momentum dependence of RIXS spectra along (0, 0)$\rightarrow$(0.44, 0) and (0, 0)$\rightarrow$(0.32, 0.32) for Par, Opt and Over. Spectra are recorded in $\pi$ polarization at the maximum of the Fe L$_{3}$ absorption edge. We show experimental data (green dots), background and elastic (purple dotted line) and  magnetic peaks (black dotted line). The sum of background, elastic and magnetic peaks is depicted as red solid line. At low q$_{//}$ a fitting is unreliable, so no fitting was attempted.}
\end{figure}

For the Par sample at (0.44, 0), a clear peak is observed at about 150 meV as shown in Figs.\ref{fig:fig2}a and b. We have tracked down this peak as a function of q and incoming beam polarization ($\sigma$ and $\pi$, as defined in Fig.\ref{fig:fig1}b). Fig.\ref{fig:fig2}c shows the strong dispersive nature of the peak along the (1, 0) direction with the peak position displaying maximum energy at high q transferred and decreasing moving towards the $\Gamma$ point, where it merges to the elastic line and is no longer resolved. A similar pattern is observed also along (0, 0)$\rightarrow$ (0.32, 0.32) (as shown in Fig.\ref{fig:fig3}). This dispersive mode is ascribed to spin excitations in line with what is observed by INS on parent NaFeAs \cite{zhang_effect_2014}. From polarization studies we observed that an incoming beam with $\pi$ polarization maximizes the spin excitations while minimizing the elastic line (Fig.\ref{fig:fig2}c). The reduction of the elastic channel in $\pi$ polarization is known in scattering theory \cite{_wiley:_????} and the polarization was therefore selected accordingly. 

Considering the low $\mu$ estimated by neutron scattering experiments \cite{li_structural_2009, johnston_puzzle_2010}, the detection of magnetic excitations in Par is remarkable, highlighting the high sensitivity of RIXS to fluctuating, rather than static, magnetic moments. A phenomenological comparison with INS data from Ref.\cite{zhang_effect_2014} confirms the magnetic nature of the mode detected in Par, ruling out the appearance of sharp electron-hole pair excitations. 

In Fig.\ref{fig:fig2}a, we show a typical fitting of the full emission line. The fitting analysis has been carried out similarly to what is described in Ref. \cite{hancock_evidence_2010} employing the following formulas: 

\begin{equation}
I_{fit} = (\beta x^{2} + \alpha x + c )  \cdot(1 - g_{\gamma}) + I_{0}\exp(ax) \cdot g_{\gamma} + G \nonumber \\
\end{equation} 
\textrm{with}\\

\begin{equation}
g_{\gamma} =  \left(\exp \left(\frac{x + \omega^{\ast}} {\Gamma}\right) + 1\right)^{-1} \nonumber \\
\end{equation} 
\textrm{and}\\
\begin{equation}
G = \frac{A}{\sigma \sqrt{2\pi}} \exp\left(\frac{(x + x_{0})^2} {2\sigma^2}\right) \nonumber \\
\end{equation} 

\begin{figure}
\includegraphics[scale=0.4]{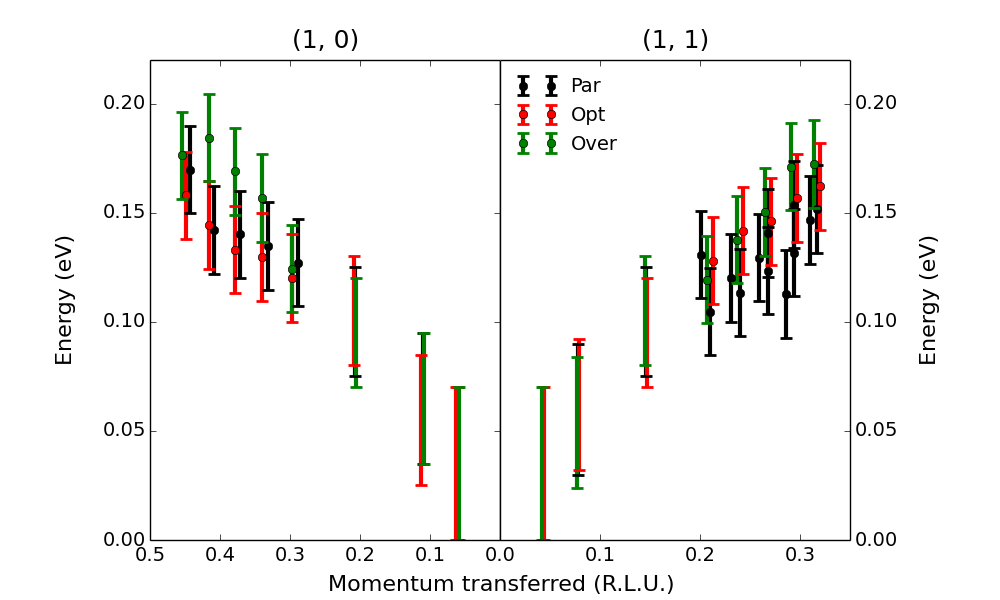}
\caption{\label{fig:fig4} Energy dispersion curves for Par, Opt and Over vs. transferred momentum. Left column: (1, 0) direction; right column: (1, 1) direction. At q = 0.06, 0.11 and 0.2 along (1, 0) and q = 0.04, 0.08 and 0.15 along (1, 1) the fitting was not possible because of overlap of the magnetic peak with the elastic line. Only an estimation of the energy range is provided for these values, as depicted by error bars. The points have been slightly shifted along the x axis for better visualization.}
\end{figure}

\begin{figure}
\includegraphics[scale=0.4]{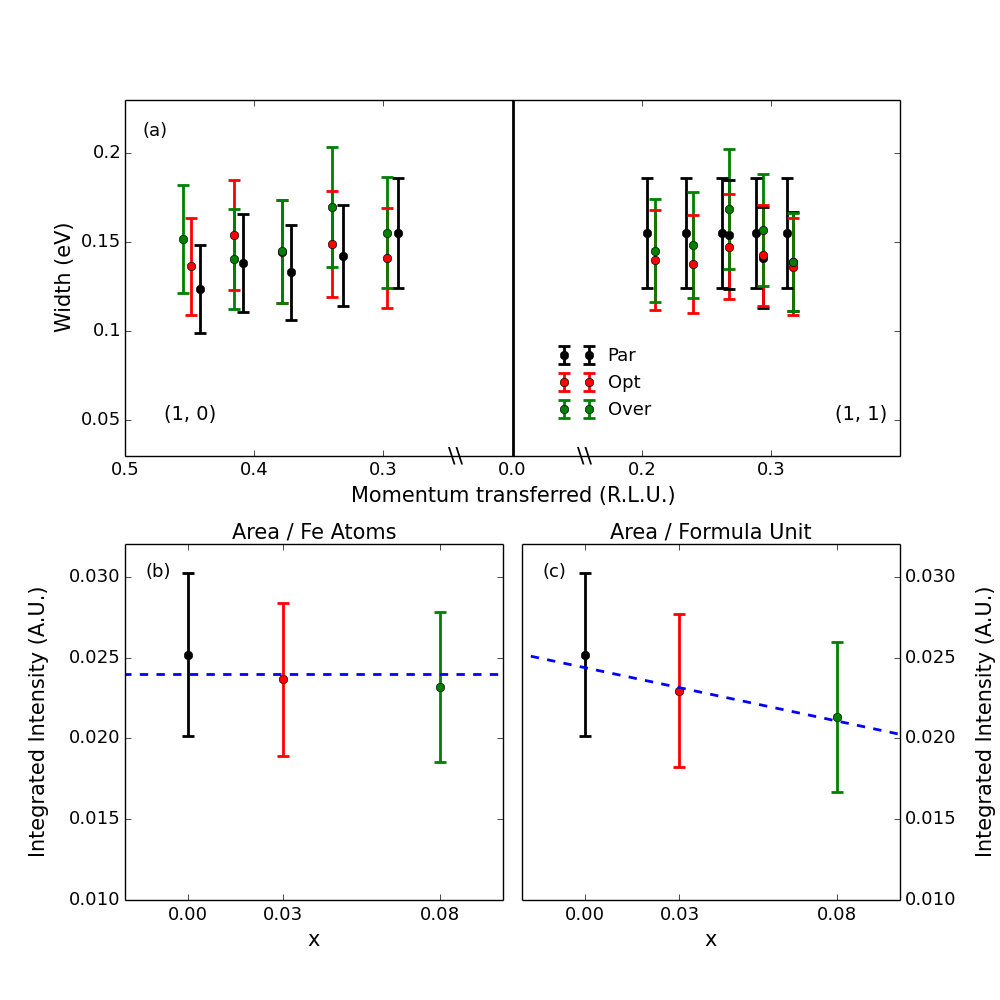}
\caption{\label{fig:fig5} (a) Fitting results. Widths of magnetic peaks vs. transferred momentum. Left side: (1, 0) direction; right side: (1, 1) direction. (b) Sum of the area of magnetic peaks for (0.44, 0), (0.41, 0), (0.37, 0), (0.32, 0.32), (0.29, 0.29) and (0.27, 0.27) normalized to Fe content vs doping (x). (c) Sum of the area of magnetic peaks for (0.44, 0), (0.41, 0), (0.37, 0), (0.32, 0.32), (0.29, 0.29) and (0.27, 0.27) rescaled per formula unit vs doping (x). These peaks were selected because they are the most clear and  with the highest intensity. The blue dotted lines represent a guide for the eye.}
\end{figure}

This approach fits the low values of energy loss (x)  with a polynomial that is changed to exponential decay at higher values of energy loss. The crossover between the two regions is obtained employing the function g$_\Gamma$, having the energy $\omega^{\ast}$ and the width $\Gamma$ as parameters \cite{hancock_evidence_2010}.
To better fit the the low energy part of the background, we employ a second order polynomial instead of the linear term proposed by Ref. \cite{hancock_evidence_2010}. This correction can be explained in the framework of the Mahan-Nozieres-De Dominicis model \cite{nozieres_threshold_1974, privalov_x-ray_2001, mahan_emission_1977, almbladh_effects_1977} as arising from many body effects happening during absorption and emission of a resonant photon in metallic systems. An additional gaussian term (introduced as G) has been introduced to fit the high energy shoulder of the main emission line observed at -4.5 eV. This spectroscopic signature was not observed in tellurides \cite{hancock_evidence_2010} but it seems to be common in pnictides as shown by Refs. \cite{zhou_persistent_2013, yang_evidence_2009}. The result of this fitting analysis is plotted in Figs.\ref{fig:fig2}a and b for Par at (0.44, 0). 

In Fig.\ref{fig:fig2}d, we show the full RIXS spectra acquired for Par, Opt and Over at (0.44, 0) with $\pi$ polarization and $h\nu=707$ eV. The main emission line, centered at -2 eV, is similar for Par, Opt and Over. This allows to employ reliably the same fitting procedure for all the samples, confirming that the background has a similar slope in all the samples. 
Fig.\ref{fig:fig3} shows the low energy range of the RIXS spectra acquired for Par, Opt and Over along (0, 0) $\rightarrow$(0.44, 0) and (0, 0)$\rightarrow$ (0.32, 0.32).
As in Ref. \cite{zhou_persistent_2013}, we employed a resolution limited gaussian curve to fit the elastic line and an anti-symmetrized lorentzian for the magnetic peak. The reason for the use of a lorentzian shape is due to the coupling of spin modes to the Stoner contiuum as observed in pnictides and doped cuprates \cite{zhou_persistent_2013, dean_persistence_2013, le_tacon_intense_2011, lee_asymmetry_2014, ishii_high-energy_2014,  peng_magnetic_2015}(see also Supplemental Material). The elastic line intensity is observed to gradually increase from Par to Opt and then to Over. We believe that this enhancement is not linked to surface or sample quality but to a real increase of diffuse scattering arising from the Fe replacement by Co. 

The bandwidth of magnetic excitations of NaFeAs is renormalized to lower values compared with AF BaFe$_{2}$As$_{2}$, as suggested by the lower T$_N$ \cite{zhou_persistent_2013, zhang_effect_2014}.  
Moving to doped samples, we discuss the main result of this work: The persistence of dispersive spin excitations in both optimal and overdoped samples (Fig.\ref{fig:fig3}). The peaks ascribed above to magnetic excitations for Par are still present at (0, 0) $\rightarrow$ (0.44, 0) and (0, 0)$\rightarrow$ (0.32, 0.32) in both doped samples as depicted in the raw data Fig.\ref{fig:fig3} and in the fitting results of Fig.\ref{fig:fig4}. Their lineshape is conserved despite large amount of Fe being substituted by Co (8 $\%$ in the overdoped sample). Furthermore, the energy of the dispersion curves does not display significant modification between the samples as shown in Fig.\ref{fig:fig4}, where the energies resulting from the fitting are summarized. Since Opt is not magnetically ordered, but superconducting, we reveal the coexistence of spin fluctuations with Cooper pairs. This entails, an at least partially, localized nature of the spins in NaFe$_{1-x}$Co$_{x}$As and a sizeable magnetic coupling in the superconducting phase. Moreover, the presence of spin excitations in the Over sample is even more astonishing than in Opt, since the high doping level is far away from AF and at the end of the superconducting dome making spin fluctuations unexpected in this region of the phase diagram. Because of high intralayer doping a possible explanation for this can be assigned to randomness and dilution of AF bonds. Similar arguments have been employed in electron doped cuprates to explain the hardening of spin excitations \cite{jia_persistent_2014}.

In Fig.\ref{fig:fig5}a we display the width of the magnetic excitations' peak. In all compounds the peaks are broader than the experimental resolution indicating a mixing of spin excitations with the Stoner continuum \cite{stanev_spin_2008}. We do not observe further broadening of magnetic peaks upon doping (Fig.\ref{fig:fig5}a). This indicates that the damping of such quasiparticles is unaffected by modification of Fermi surface topology, chemical potential and scattering rate caused by doping \cite{ye_extraordinary_2014}.  
To quantify the intensity, we integrated the magnetic peaks and summed the highest q points. We employed the values obtained at q$_{//}$ = (0.44, 0), (0.41, 0), (0.37, 0), (0.32, 0.32), (0.29, 0.29) and (0.27, 0.27). 
In these spectra, the peaks are well resolved from the elastic line and the intensity is maximal. Since we normalized to the Fe fluorescence line, the intensity can be interpreted as the magnetic weight per Fe atom. The results in Fig.\ref{fig:fig5}b show that the magnetic weight per Fe atom is preserved (Fig.\ref{fig:fig5}b) after doping, indicating magnetism residing on Fe atoms. Nonetheless, the Fe content decreases upon doping (Co replaces Fe), thus decreasing the absolute RIXS signal. This implies that the magnetic spectral weight has to be rescaled to the overall formula unit (multiplying by the relative factors (1-0.03) and (1-0.08) for Opt and Over, respectively). This renormalization slightly reduces the integrated intensity of spin excitations due to dilution of Fe by Co (see Fig.\ref{fig:fig5}b and c). 

\begin{figure}
\includegraphics[width=\textwidth]{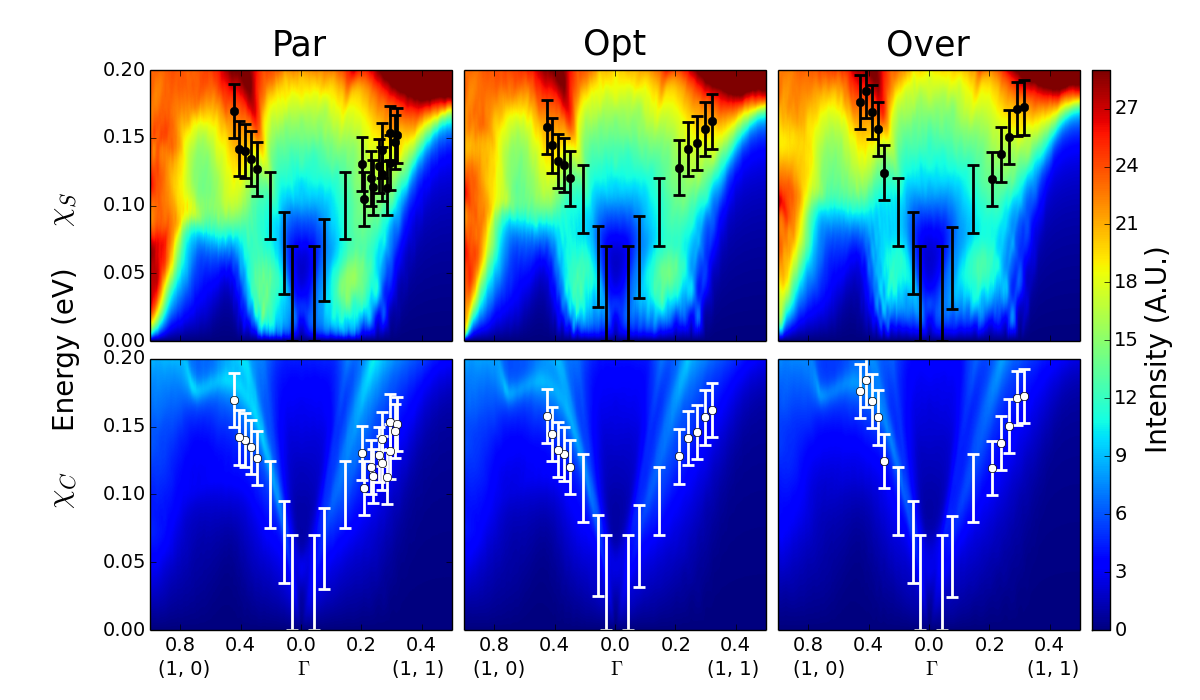}
\caption{\label{fig:fig6} Calculated dynamical susceptibility (color plot) overlaid with experimental data (black and white dots with error bars) for Par, Opt and Over. Top row: Calculated spin susceptibility ($\chi_{S}$). Bottom row: Calculated charge susceptibility ($\chi_{C}$). Both crystallographic directions (1, 0) and (1, 1) are shown.} 
\end{figure}

To further elucidate the experimental data, we calculated the dynamical susceptibility for all the samples.
The self-energy dressed $\chi_{S}$ and $\chi_{C}$ were disentangled as displayed in the color plot in Fig.\ref{fig:fig6}. Our calculations show that the intensity of $\chi_{S}$ is almost an order of magnitude higher than $\chi_{C}$, and the latter is pushed slightly higher in energy than the former. This is due to electronic correlations that shift the spectral weight of charge modes to higher energies leaving the spin part to reside at lower energies. Comparison of $\chi_{S}$ along the (1, 0) and (1, 1) directions clearly reveals that the spin modes are more pronounced at (1, 0). The large $\chi_{S}$ around (1, 0) is the residual of AF ordering, indicating the ability  of theory to catch the tendency towards AF ordering with underdoping. As RIXS is sensitive to the dynamical susceptibility \cite{ament_resonant_2011}, we confirm that the main spectroscopic weight detected in the low energy region ($<$200 meV) after background subtraction in NaFe$_{1-x}$Co$_{x}$As is mainly of magnetic origin. This comparison between experiments and calculations is carried out on a qualitative basis because calculations involving matrix elements effects and the presence of a core hole in the intermediate state of the RIXS process should be included to quantitatively compare $\chi$ calculations with the experimental RIXS intensity.
Our calculations for the same Co doping levels as measured display that doping slightly shifts the spin modes upwards in energy. 
Comparing experimental data (dots in Fig.\ref{fig:fig6}) and calculations (color plot in Fig.\ref{fig:fig6}) we find that the extracted dispersions follow the highest intensity region of the calculated $\chi_{S}$. The small hardening of $\chi_{S}$ indicated by calculations has not been observed in our RIXS measurements, likely being beyond the resolving capacity of our current instrumentation. The decrease of spectral weight for increasing Co doping observed in our experiment is qualitatively captured by the calculations, as can be observed by the  intensity colorscale of Fig.\ref{fig:fig6}. However, this comparison is not meant to be quantitative since in the calculations the matrix elements have not been taken into account.

The phenomenological agreement of the energy of spin excitations in theory and experiment confirms the robustness of spin excitations in NaFe$_{1-x}$Co$_{x}$As even when a large amount of Fe has been substituted by Co. In high $\mu$ FePns, such as the BaFe$_{2}$As$_{2}$ series, the energy of magnetic correlations for BaFe$_{2-x}$Ni$_{x}$As$_{2}$ and Ba$_{0.6}$K$_{0.4}$Fe$_{2}$As$_{2}$ is, relatively to the parent compound, unchanged and softened, respectively \cite{luo_electron_2013, wang_doping_2013, zhou_persistent_2013}. This illustrates that the evolution of spin fluctuations is affected by doping site (inside or outside of the Fe layer) and type of carriers (electrons or holes). These observations demonstrate that the presence of a sizeable magnetic coupling seems to be universal, whereas its evolution with doping is not. 

\section{\label{sec:level1}Conclusions}
In conclusion, we have performed RIXS measurements and DFT-based calculations of dynamic susceptibilities on low static $\mu$ NaFe$_{1-x}$Co$_x$As (x = 0, 0.03 and 0.08). We observe broad and dispersive spin excitations around 150 meV from (0, 0)$\rightarrow$ (0, 0.44) and (0, 0)$\rightarrow$ (0.32, 0.32) in all the samples measured. In AF NaFeAs the spin excitations, remarkably, manifest themselves despite its low $\mu$. This confirms the quantum fluctuating nature of spins in NaFeAs and demonstrates the ability of RIXS to probe spin correlations in low $\mu$ itinerant systems. In the optimally doped samples, magnetic ordering is replaced by SC with spin correlations surviving and preserving their bandwidth. Measurements on the overdoped compound reveal the presence of magnetic modes also when SC has been suppressed and a metallic phase has taken over. The spectral weight of such modes seems to gradually decrease with Co doping.
Our experiments demonstrate that the suppression of SC is not linked with a complete disappearance of high energy magnetic excitations, as magnetic coupling is present also in overdoped samples, but is rather connected to subtle effects likely happening at lower energy scale.

\begin{acknowledgments}
J.P. and T.S. acknowledge financial support through the Dysenos AG by Kabelwerke Brugg AG Holding, Fachhochschule Nordwestschweiz, and the Paul Scherrer Institut. Experiments have been performed at the ADRESS beamline of the Swiss Light Source at Paul Scherrer Institut. Part of this research has been funded by the Swiss National Science Foundation through the Sinergia network Mott Physics Beyond the Heisenberg (MPBH) model and the D-A-CH programme (SNSF Research Grant 200021L 141325). The work at IOP-CAS is supported by NSF \& MOST through research projects.
\end{acknowledgments}

\end{document}